\documentclass[utf8]{aa} 

\usepackage{graphicx}
\usepackage{amsmath,amsfonts,amssymb}
\usepackage{txfonts}
\usepackage{natbib} 
\bibpunct{(}{)}{;}{a}{}{,} 

\usepackage{verbatim}
\usepackage{url}
\usepackage{epstopdf}

\begin{document}

\title{Mixing-length estimates from binary systems. A theoretical investigation on the estimation errors} 

\subtitle{}

\author{G. Valle \inst{1,2}, M. Dell'Omodarme \inst{1}, P.G. Prada Moroni
	\inst{1,2}, S. Degl'Innocenti \inst{1,2} 
}
\titlerunning{Mixing-length from binary systems}
\authorrunning{Valle, G. et al.}

\institute{
	Dipartimento di Fisica "Enrico Fermi'',
	Universit\`a di Pisa, Largo Pontecorvo 3, I-56127, Pisa, Italy
	\and
	INFN,
	Sezione di Pisa, Largo Pontecorvo 3, I-56127, Pisa, Italy
}

\offprints{G. Valle, valle@df.unipi.it}

\date{}

\abstract
{
We performed a theoretical investigation on the biases and random uncertainties affecting the recovery of the mixing-length parameter $\alpha_{\rm ml}$ from an ideal eclipsing double-lined binary system, with well constrained masses and radii.
We focused on a test case composed by a primary of mass $M = 0.95$ $M_\odot$ and a secondary of  $M = 0.85$ $M_\odot$. Synthetic stars were generated coeval and with a common chemical composition by sampling from a dense grid of stellar models. Observational errors were simulated by adding random perturbations to mock data. The $\alpha_{\rm ml}$ parameter was then recovered by means of the SCEPtER-binary pipeline.
Several Monte Carlo simulations were conducted considering three metallicities, coupled to three different evolutionary stages of the primary. For each configuration artificial data were sampled assuming an increasing difference between the mixing-length of the two stars.
The mixing length values were then reconstructed adopting three alternative set-ups. A first method, which assumes full independence between the two stars, showed a great difficulty to constrain the mixing-length values: the recovered values were nearly unconstrained with a standard deviation of about 0.40.
The second technique imposes the constraint of common age and initial chemical composition for the two stars in the fit. We found that $\alpha_{\rm ml,1}$ values closely match the ones  recovered under the previous configuration, but $\alpha_{\rm ml,2}$ values are much more peaked around  unbiased estimates. This occurs because the primary star provides a much more tight age constraint in the joint fit than the secondary, thus leading to the rejection of several extreme solutions for the secondary. Within this second scenario we also explored, for systems sharing a common $\alpha_{\rm ml} = 2.0$, the difference in the mixing-length values of the two stars only due to random fluctuations owing to the observational errors. The posterior distribution of these differences was peaked around zero, with a somewhat large standard deviation of 0.3 (about 15\% of the solar-scaled value). Therefore about 32\% of systems with true identical $\alpha_{\rm ml}$ are expected to show differences higher than that only owing to random errors. 
The third technique also imposes the constraint of a common mixing-length value for the two stars. This assumption is generally not true for the sample stars, and served as a test  for identification of wrong fitting assumptions. In this case the common mixing-length is mainly dictated by the value of $\alpha_{\rm ml,2}$. However, an increasing share of systems cannot be fitted by the algorithm as the differences of $\alpha_{\rm ml}$ between the two stars in the synthetic systems increases. For $\Delta \alpha_{\rm ml} > 0.4$ less than half of the systems can be recovered and only 20\% at $\Delta \alpha_{\rm ml} = 1.0$. }


\maketitle

\section{Introduction}\label{sec:intro}

Despite the huge refinements in the accuracy and reliability of 
the stellar evolutionary predictions, several mechanisms involved in the evolution of stars are still poorly understood. A major and long standing problem affecting stellar model computations is the treatment of superadiabatic convection. This lack prevents a firm and reliable prediction of the extension of the external convective regions.

A precise treatment of external convection would require 3D hydrodynamical calculations, greatly improved in recent years \citep[see e.g.][]{Tanner2013, Trampedach2014, Magic2015} however they still cannot cover the wide range of input physics needed to model stellar computations. Moreover defining alpha in one dimensional codes from the results of 3D simulations is quite ambiguous, due to the differences between the mixing length representation of convection and the convection behaviour in 2D/3D simulations, even if there have been relevant attempts in literature \citep{Lydon1992, Ludwig1999, Tanner2013, Magic2015, Mosumgaard2017}.
Therefore the current generation of stellar
evolution codes still address this problem by relying, almost universally, on the mixing-length theory \citep{bohmvitense58}.
In this framework the efficiency of the convective transport and
the stellar structure in the superadiabatic transition layers 
depends on the mixing-length $l$, 
which is supposed to be proportional to the pressure scale height $H_p$,
i.e. $l= \alpha_{\rm ml} H_p$, where $\alpha_{\rm ml}$ is a non-dimensional free 
parameter to be somehow calibrated. 

As a result of this freedom, neither the
effective temperature nor the radius of stars with a thick outer
convective envelope can be firmly predicted  by current generation of 1D stellar models since they strongly
depend on the calibrated value of $\alpha_{\rm ml}$. This obviously influences the stellar characteristics recovered by fit techniques that exploit these observables.

The classical target for mixing-length calibration is the Sun, but
the generalization of this calibration to different evolutionary phases,
metallicity, and mass ranges has been questioned both on
theoretical and observational grounds. 
In particular, a growing amount of observations suggests that the adoption of
the solar calibrated $\alpha_{\rm ml}$ does not allow to properly model all kind of
stars \citep[see e.g][]{Guenther2000, Yildiz2006, Yildiz2007, Clausen2009, Deheuvels2011, Bonaca2012, Mathur2012, Wu2015, Joyce2018,Joyce2018b, Li2018}.

Several of the above mentioned studies consider stars in double-lined detached eclipsing binary systems. In fact, in this case it is possible to obtain an accurate measurement of both masses and radii of the two stars. The availability of these two fundamental quantities allows for a stringent tests of the stellar models \citep[see e.g.][]{Claret2007, Stancliffe2015, Gallenne2015, Claret2016, TZFor, Claret2017} and thus also on the possible variations of the mixing-length parameter in different mass and metallicity ranges.

A recent theoretical analysis \citep{ML-campo} pointed out many limitations for a mixing-length calibration from field stars, so it is interesting to investigate the question of the reliability of this calibration in binary systems. The aim of this work is to perform such investigation for a system composed by two low mass stars, thus avoiding the supplementary complication of dealing with the concurrent calibration of the convective core overshooting parameter.

The investigation is focussed on a synthetic binary system composed by a primary of $M_1 = 0.95$ $M_{\odot}$ and a secondary of $M_2 = 0.85$ $M_{\odot}$, resulting in a mass ratio of about 1.1. This configuration, quite common for real systems, was chosen also to allow the stars to be sampled in different evolutionary phases, because this provides the most stringent constraints for the recovery \citep{TZFor, Claret2016}. The focus of our analysis is in quantifying the errors in the recovered mixing-length values
arising only from observational errors. Therefore no systematic discrepancies between the grid of models adopted in the recovery and the artificial stars are assumed. Although this is a far too optimistic assumption when dealing with real world binary systems, nevertheless it is a mandatory step that allows to asses the very minimum errors that can be expected in the calibration process. Moreover only a theoretical investigation can highlight the presence of hidden biases and dependencies \citep[see e.g.][for an analysis of spurious metallicity dependencies of the recovered mixing-length on field stars]{ML-campo} that would be otherwise neglected.

\section{Methods}\label{sec:method}

The analysis was performed at three different metallicity values, centred around the solar value ($Z$ = 0.0074, 0.0129, 0.0221), and at three different evolutionary stages of the primary star. More precisely, we define a relative age $r$ with respect to the main sequence (MS) lifetime and we selected models at $r$ = 60\%, 90\%, and 120\% of the central hydrogen exhaustion time. The first two points correspond to a primary at the middle and nearly at the end of the MS, while the third one corresponds to a primary in the red giant branch (RGB) phase. Correspondingly, the secondary star is always in the MS, at about 40\%, 55\%, and 75\% of its central hydrogen exhaustion time.
The synthetic stars' initial helium abundances were obtained by the linear relation $Y = Y_p+\frac{\Delta Y}{\Delta Z} Z$ with the primordial abundance $Y_p = 0.2485$ from WMAP
\citep{peimbert07a,peimbert07b} and with a helium-to-metal enrichment ratio $\Delta Y/\Delta Z = 2.0$ \citep[see e.g.][]{gennaro10}.

We considered different possible values of $\alpha_{\rm ml}$ for the two stars. The reference scenario has the two stars with common mixing-length parameter $\alpha_{\rm ml,1} = \alpha_{\rm ml,2} = 2.0$. The other five cases adopted a systematic increasing difference between of the two mixing length values in step of 0.2: while the mixing-length of the primary star was increased in step of 0.1 ($\alpha_{\rm ml,1}$ = 2.1, 2.2, 2.3, 2.4, and 2.5), the values of  $\alpha_{\rm ml,2}$ were correspondingly decreased by 0.1 ($\alpha_{\rm ml,2}$ = 1.9, 1.8, 1.7, 1.6, 1.5). Thus the extreme scenario has a difference in mixing-length value $\alpha_{\rm ml,1}-\alpha_{\rm ml,2}$ = 1.0.
The considered simulation parameters are summarized in Table~\ref{tab:summary}.
 
\begin{table*}[ht]
	\caption{Summary of the simulation parameters for the considered synthetic binary systems.} \label{tab:summary}
	\centering
	\begin{tabular}{lcc}
		\hline\hline
		Parameter & Primary & Secondary\\
		\hline
		Mass ($M_{\odot}$) & 0.95 & 0.85 \\ 
		Relative ages & 0.60, 0.90, 1.20 & 0.38, 0.57, 0.76\\
		$\alpha_{\rm ml}$ & 2.0, 2.1, 2.2, 2.3, 2.4, 2.5 & 2.0, 1.9, 1.8, 1.7, 1.6, 1.5 \\
	    $Z$ & 0.0074, 0.0129, 0.0221 & 0.0074, 0.0129, 0.0221 \\
	    $\Delta Y/\Delta Z$ & 2.0 & 2.0 \\ 
		\hline
	\end{tabular}
\end{table*}

For each possible mixing length combination, metallicity, and evolutionary stage artificial stars were sampled from the grid described in Sect.~\ref{sec:grids}, and their observables perturbed by means of a Monte Carlo procedure assuming Gaussian errors. The procedure was repeated $N = 5000$ times for each artificial system. The adopted observable constraints were the effective temperature $T_{\rm eff}$, metallicity [Fe/H], mass $M$, and radius $R$ of both stars. We assumed uncertainties of 100 K in $T_{\rm eff}$, 0.1 dex in [Fe/H], 0.5\% in $M$ and 0.25\% in $R$. Similar precisions on the mass and radii are achievable for a small subset of binary systems, but they are mandatory for calibration purposes \citep[see e.g.][]{TZFor}. 

\subsection{Grid-based recovery technique}\label{sec:method-gd}

Each artificial system was then reconstructed adopting the SCEPtER-binary pipeline \citep{binary}, modified to consider the mixing-length value in the estimation process. Details about the technique can be found in \citet{binary, testW}.

We adopted the pipeline in three configurations.
The first one fits the two stars independently, thus effectively losing the binary constraint. This result is useful as a reference for comparison with the other results. A second configuration imposes the 
binary constraint, forcing the pipeline to return identical ages, initial metallicity $Z$, and initial helium abundance $Y$ for the two stars; however each star can have different mixing-length values. The third configuration adds the constraint of a common mixing-length value. The latter scenario is particularly useful to test the sensitivity of grid techniques in identifying an inadequate fitting model specification. Indeed, as the difference in the mixing length of the synthetic stars grows, the constraint of a common $\alpha_{\rm ml}$ value would become more and more difficult to satisfy, causing several fits to return no acceptable values for the system. The adopted procedure will allow to quantify the theoretical fraction of systems for which such behaviour is expected.

\subsection{Stellar models grid}
\label{sec:grids}

The estimation procedure required a grid of stellar models, sufficiently extended to cover the whole parameter space. To this purpose, we adopted the same stellar models grid as in \citet{ML-campo}, 
 computed by means of the FRANEC code \citep{scilla2008, Tognelli2011}, in the same
configuration as was adopted to compute the Pisa Stellar
Evolution Data Base\footnote{\url{http://astro.df.unipi.it/stellar-models/}} 
for low-mass stars \citep{database2012}. Models were calculated for the solar heavy-element mixture by \citet{AGSS09}. 
Atomic diffusion was included, taking into account the
effects of gravitational settling and thermal diffusion with
coefficients given by \citet{thoul94}. 
Outer boundary conditions were determined by integrating the $T(\tau)$ relation by \citet{KrishnaSwamy1966}.
Further details on the stellar models can be found in \citet{cefeidi,eta,binary} and references therein.  Although the choices in the input physics play a relevant role when estimating stellar parameters from real observational data, they are of minor relevance for our aim because artificial stars are recovered from the same model grid adopted for their sampling.

\begin{figure}
	\centering
	\includegraphics[height=8cm,angle=-90]{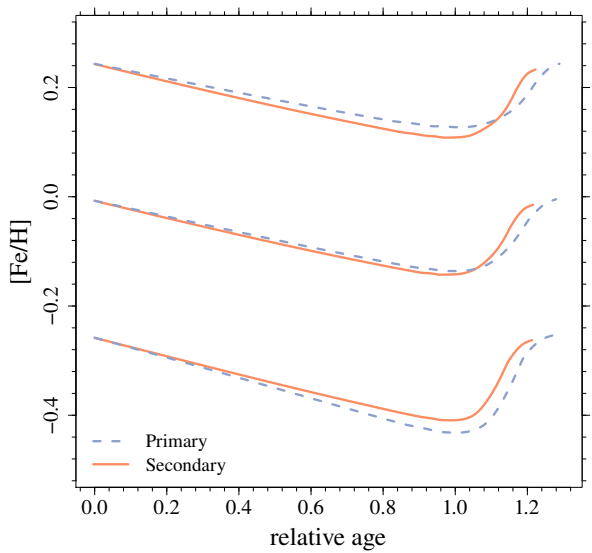}
	\caption{Evolution of the surface [Fe/H] for primary (dashed lines) and secondary (solid lines) stars as a function of the stellar relative age (relative age is 0.0 at ZAMS and 1.0 at the end of the MS). Models for the three assumed initial metallicities are represented.}
	\label{fig:feh}
\end{figure}

The adoption of microscopic diffusion in the stellar computations causes an evolution with time of the surface [Fe/H], which is adopted as one of the observational constraints in the analysis. Figure~\ref{fig:feh} shows the trend of the surface [Fe/H] as a function of the relative age. As well known, the [Fe/H] abundance reaches a minimum around the end of the MS, while increases at later time due to the first dredge-up.

The grid of models spans the range [0.8, 1.0] $M_{\odot}$, with a step of 0.01 $M_{\odot}$, and covers  the 
initial metallicity interval $-0.4 \; {\rm dex} \leq$ [Fe/H] $\leq 0.4$ dex, with
a step of 0.05 dex.  
For each metallicity we computed models for nine different values of the initial helium abundance by following the above mentioned 
linear relation $Y = Y_p+\frac{\Delta Y}{\Delta Z} Z$,
with a helium-to-metal enrichment ratio $\Delta Y/\Delta Z$
from 1 to 3 with a step of 0.25 \citep{gennaro10}. $\Delta Y/\Delta Z = 2.0$ corresponds to the reference value for the synthetic systems. Ultimately, the grid spans a set of 153 different initial chemical compositions. For each mass, metallicity and initial helium abundance, we computed models for
 21 values of the mixing-length parameter $\alpha_{\rm ml}$ in the range
[1.0, 3.0] with a step of 0.1. With the assumed input physics, the solar-calibrated value  is $\alpha_{\rm ml} = 2.1$. All the adopted steps are sufficiently small to impact in a negligible way on the estimates.

\section{Results}
\label{sec:results}

The analysis of the fit results for the considered scenarios revealed some expected behaviours and also some peculiar effects. The following subsections explore in detail the outcomes in the three considered fitting configurations. 

\subsection{Independent recovery}\label{sec:independent}

\begin{figure*}
	\centering
	\includegraphics[height=16.5cm,angle=-90]{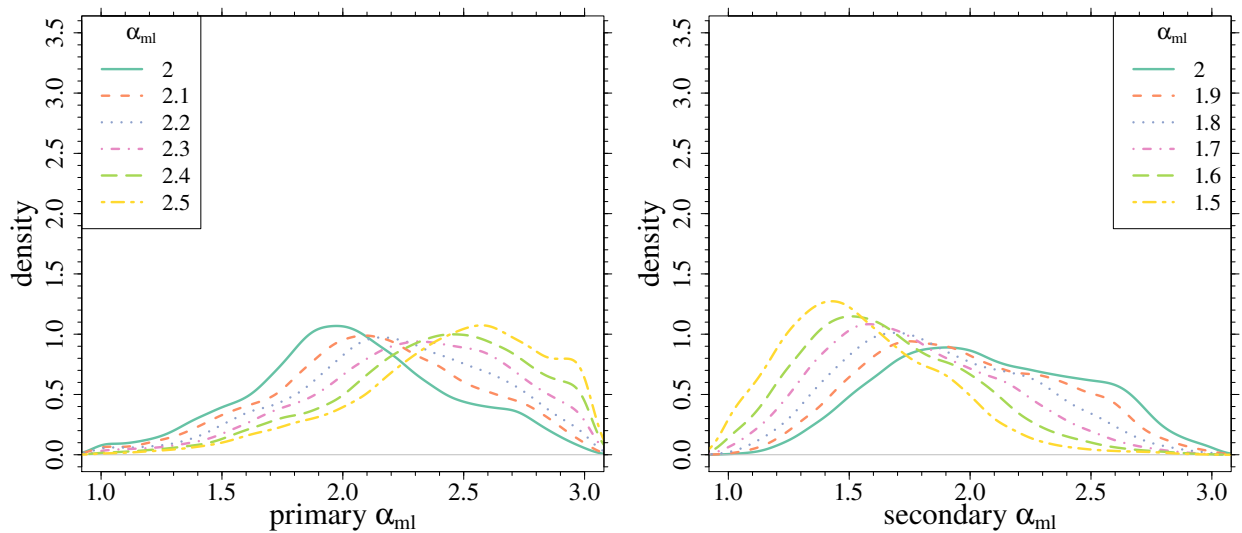}
	\caption{({\bf Left}): density of probability of mixing-length value estimates for the primary
		star, for the independent stellar fit. Different line styles identify the $\alpha_{\rm ml}$ values from (2.0 to 2.5) adopted in the
		sampling of the mock data. ({\bf Right}): same as in the left panel, but for the secondary star. The sampling $\alpha_{\rm ml}$ values for the secondary star run from 2.0 to 1.5.}
	\label{fig:independent}
\end{figure*}

The fit of the binary systems under full independence between the stars (i.e. stars can be fitted at different chemical compositions and different ages) revealed a great difficulty to constrain the mixing-length value for the two stars. The recovered marginalized posterior density of $\alpha_{\rm ml,1}$ and $\alpha_{\rm ml,2}$ are presented in Fig.~\ref{fig:independent} and Table~\ref{tab:independent}. While the mean values of the mixing-length is in general consistently recovered for both stars, there is a huge variability in the results, which practically cover the whole allowed $\alpha_{\rm ml}$ range. 
The $\alpha_{\rm ml,1}$ values are underestimated for the two most extreme scenarios ($\alpha_{\rm ml,1}$ = 2.4, 2.5), as a consequence of an edge effect that truncates the estimates at the grid upper value ($\alpha_{\rm ml}$ = 3.0). This effect is clearly evidenced in Fig.~\ref{fig:independent}A: for the sampling at $\alpha_{\rm ml}$ = 2.5 the posterior density is clearly truncated at the upper edge.
A correspondent tendency to overestimation is reported in the last two cases for the secondary star ($\alpha_{\rm ml,2}$ = 1.6 and 1.5).
These results confirm the theoretical finding by \citet{ML-campo}, obtained for field stars. It seems that even from stars with exceptionally well constrained masses and radii, a calibration of the mixing-length parameter is poorly reliable. 

\begin{table}[ht]
	\caption{Mean and standard deviation of the recovered mixing-length values for primary and secondary stars, in dependence on the $\Delta \alpha_{\rm ml}$ adopted in the generation. The reference $\alpha_{\rm ml}$ of synthetic data are listed in the second and third columns.} 
\label{tab:independent}
	\centering
	\begin{tabular}{lll|cc|cc}
		\hline\hline
		& & & \multicolumn{2}{c|}{Primary} & \multicolumn{2}{c}{Secondary}\\
		$\Delta \alpha_{\rm ml}$ & $\alpha_{\rm ml,1}$ & $\alpha_{\rm ml,2}$ & Mean & Sd & Mean & Sd \\ 
		\hline
		0.0 & 2.0 & 2.0 & 2.03 & 0.41 & 2.06 & 0.40 \\ 
		0.2 & 2.1 & 1.9 & 2.12 & 0.41 & 1.97 & 0.39 \\ 
		0.4 & 2.2 & 1.8 & 2.20 & 0.41 & 1.87 & 0.38 \\ 
		0.6 & 2.3 & 1.7 & 2.28 & 0.41 & 1.77 & 0.36 \\ 
		0.8 & 2.4 & 1.6 & 2.36 & 0.40 & 1.67 & 0.35 \\ 
		1.0 & 2.5 & 1.5 & 2.44 & 0.39 & 1.56 & 0.32 \\ 
		\hline
	\end{tabular}
\end{table}

\subsection{Binary constraints in age and chemical composition}\label{sec:partially-dep}

\begin{figure*}
  \begin{center}
    \includegraphics[height=16.5cm,angle=-90]{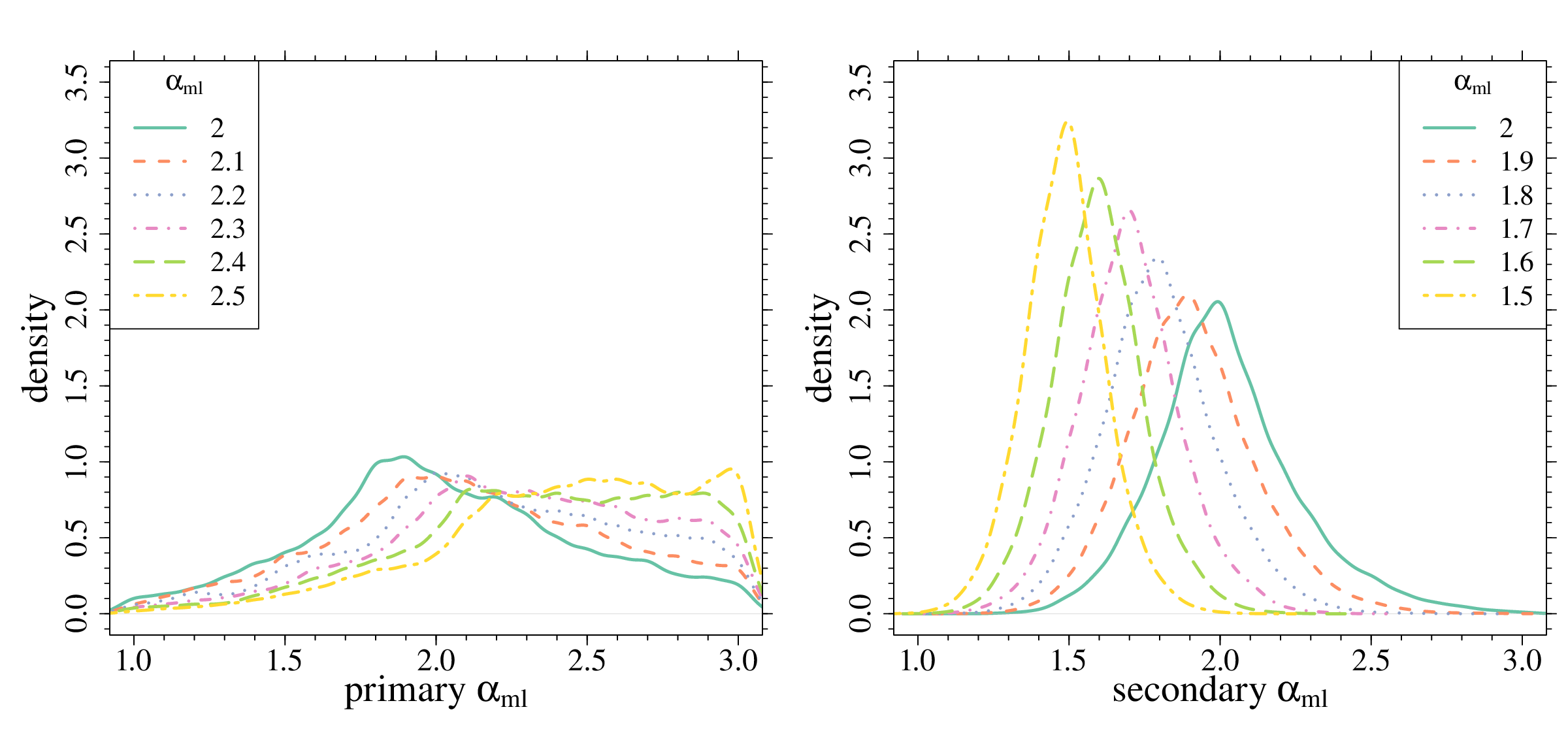}
    \end{center}
\caption{As in Fig.~\ref{fig:independent}, but assuming common age, and common initial $Z$ and $Y$ in the fit.}\label{fig:dep-no-alpha}
\end{figure*}

Imposing the constraint of a common age and initial chemical composition for the two stars modifies the results in an interesting way. As it appears from Fig.~\ref{fig:dep-no-alpha} and Table~\ref{tab:dep-no-alpha}, the recovered $\alpha_{\rm ml,1}$ values closely match the ones shown in Sect.~\ref{sec:independent}, showing the same mean values and the characteristic  large dispersion. On the other hands, the recovered $\alpha_{\rm ml,2}$ values are much more peaked around their mean values, which provide unbiased estimates of the values adopted in the sampling. These results suggest that while the mixing-length parameter of the primary star is not further constrained by the fit, this is not the case for that of the secondary. Indeed the standard deviations of the recovered $\alpha_{\rm ml,2}$ values range from one half to one third of those of $\alpha_{\rm ml,1}$. Moreover, the tendency to overestimate $\alpha_{\rm ml,2}$ for the two most extreme scenarios -- shown in Sect.~\ref{sec:independent} -- disappears, as a consequence of the much smaller variability which prevents edge effects to play a role.    

\begin{table}[ht]
	\caption{As in Table~\ref{tab:independent}, but imposing the constraints of common age, and common initial $Z$ and $Y$ for the binary components.} 
	\label{tab:dep-no-alpha}
	\centering
	\begin{tabular}{lcc|cc}
		\hline\hline
		& \multicolumn{2}{c|}{Primary} & \multicolumn{2}{c}{Secondary}\\
		$\Delta \alpha_{\rm ml}$ & Mean & Sd & Mean & Sd \\ 
		\hline
		0.0 & 2.03 & 0.44 & 2.02 & 0.24 \\ 
		0.2 & 2.11 & 0.46 & 1.91 & 0.21 \\ 
		0.4 & 2.19 & 0.46 & 1.80 & 0.19 \\ 
		0.6 & 2.28 & 0.44 & 1.70 & 0.16 \\ 
		0.8 & 2.35 & 0.44 & 1.60 & 0.15 \\ 
		1.0 & 2.43 & 0.42 & 1.49 & 0.13 \\ 
		\hline
	\end{tabular}
\end{table}

\begin{figure}
	\centering
	\includegraphics[height=8.2cm,angle=-90]{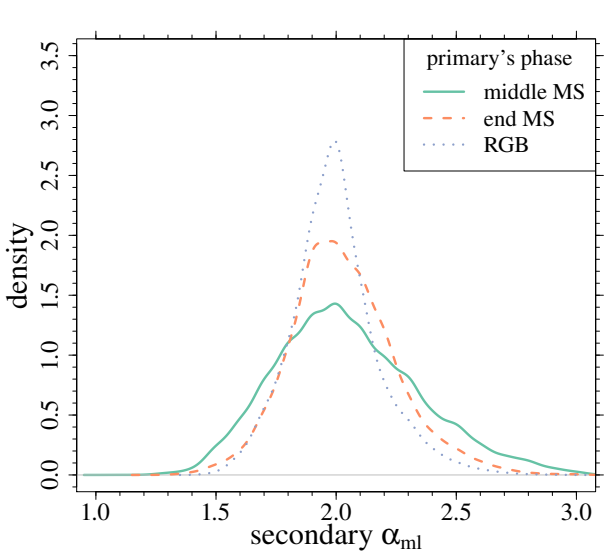}
	\caption{Density of probability of mixing-length value estimates for the secondary star for synthetic data with $\alpha_{\rm ml}$ = 2.0, as a function of the evolutionary phase of the primary star.}
	\label{fig:alpha-ML2-fase}
\end{figure}

The impressing differences in the two stars behaviour is dictated by the different constraints they provide each other in the joint fit. While the independently recovered initial chemical composition are nearly identical for both stars, this is not the case for the age. As discussed in  detail in \citet{binary}, the primary star provides a much more tight age constraint in the joint fit than the secondary. This occurs because the age relative error becomes smaller as a star evolves in the MS, due to the faster evolutionary time scale. In fact the age range allowed by the observable constraint errors is smaller in rapid evolutionary phases. This is clearly demonstrated in Fig.~\ref{fig:alpha-ML2-fase}, which presents the $\alpha_{\rm ml,2}$ posterior density for a mock data with  $\alpha_{\rm ml,1} = \alpha_{\rm ml,2}$ = 2.0, in dependence of the evolutionary stage of the primary star. The distribution of the recovered secondary mixing-length values shrinks as the primary evolves. The RGB scenario -- for which the evolutionary time scale is the fastest --  provides the lowest variance for the estimated $\alpha_{\rm ml,2}$ values.  
This effect is further shown in Table~\ref{tab:dep-no-alpha-phase}. It is apparent that the standard deviation of the recovered mixing-length value for the secondary star shrinks as the primary evolves, mainly when the differences in the sampled $\alpha_{\rm ml}$ value are lower than about 0.6. 
 
\begin{table}[ht]
	\caption{As in Table~\ref{tab:dep-no-alpha}, but splitting the result according to the evolutionary phase of the primary star. The relative age $r$ assumes value 0.0 at ZAMS and 1.0 at the end of the MS. } 
	\label{tab:dep-no-alpha-phase}
	\centering
	\begin{tabular}{ll|cccccc}
		\hline\hline
		$\Delta \alpha_{\rm ml}$ & $\alpha_{\rm ml}$ & Mean & Sd & Mean & Sd & Mean & Sd\\ 
		\hline
		& & \multicolumn{6}{c}{Primary}\\
		& & \multicolumn{2}{c}{$r$ = 0.6} & \multicolumn{2}{c}{$r$ = 0.9} & \multicolumn{2}{c}{$r$ = 1.2}\\
		\hline
		0.0 & 2.0 & 2.08 & 0.42 & 2.08 & 0.36 & 1.93 & 0.52 \\ 
		0.2 & 2.1 & 2.16 & 0.41 & 2.17 & 0.36 & 2.00 & 0.55 \\ 
		0.4 & 2.2 & 2.24 & 0.41 & 2.25 & 0.36 & 2.08 & 0.56 \\ 
		0.6 & 2.3 & 2.31 & 0.39 & 2.35 & 0.36 & 2.17 & 0.54 \\ 
		0.8 & 2.4 & 2.39 & 0.38 & 2.42 & 0.35 & 2.23 & 0.54 \\ 
		1.0 & 2.5 & 2.45 & 0.37 & 2.49 & 0.34 & 2.34 & 0.52 \\ 
		\hline
		& & \multicolumn{6}{c}{Secondary}\\
		\hline
		0.0 & 2.0 & 2.05 & 0.30 & 2.03 & 0.22 & 2.00 & 0.18 \\ 
		0.2 & 1.9 & 1.92 & 0.26 & 1.91 & 0.19 & 1.89 & 0.17 \\ 
		0.4 & 1.8 & 1.82 & 0.22 & 1.80 & 0.17 & 1.79 & 0.16 \\ 
		0.6 & 1.7 & 1.69 & 0.19 & 1.71 & 0.16 & 1.71 & 0.14 \\ 
		0.8 & 1.6 & 1.59 & 0.17 & 1.60 & 0.14 & 1.60 & 0.13 \\ 
		1.0 & 1.5 & 1.47 & 0.14 & 1.49 & 0.12 & 1.50 & 0.12 \\ 
		\hline
	\end{tabular}
\end{table}

As a consequence, the common age constraint leads to the rejection of several extreme solutions for the secondary, while the solutions of the primary are unaffected. The net result is a selection on the mixing-length values of the secondary star, which disfavours extreme variations with respect to the sampling values.

\subsubsection{Expected difference in $\alpha_{\rm ml}$ value for the two stars}

The analysis conducted assuming independent mixing-length values for the two stars also allows to estimate an interesting parameter that is, the expected dispersion of the recovered $\alpha_{\rm ml}$ values when sampled at common $\alpha_{\rm ml}$ = 2.0. An estimate of this value can help in judging how reliable are calibrations from binary systems allowing for independent mixing-length values.  

\begin{figure}
	\centering
	\includegraphics[height=8.2cm,angle=-90]{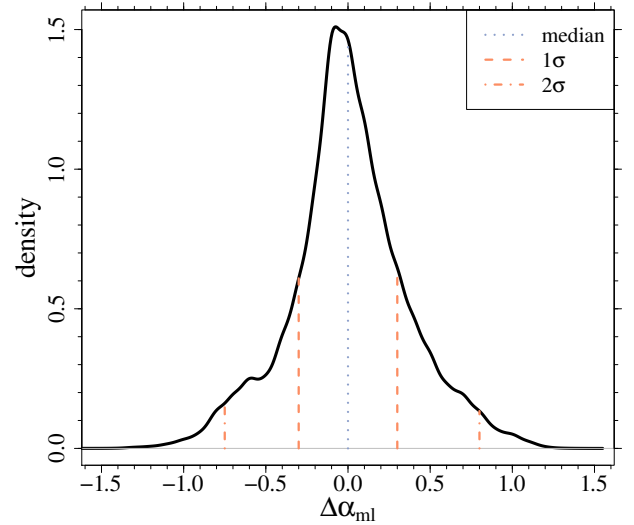}
	\caption{Distribution of the expected differences between the recovered mixing-length values of the primary and secondary star, when both stars in the synthetic systems have common $\alpha_{\rm ml}$ = 2.0.}
	\label{fig:density-diff}
\end{figure}

The question requires to consider the reconstructed differences $\Delta \alpha_{\rm ml} = \alpha_{\rm ml,1} - \alpha_{\rm ml,2}$ for all the binary systems simulated with the same $\alpha_{\rm ml}$. The distribution of these differences are only due to random errors on the observables and should be therefore considered as the minimum variability on the mixing-length parameter. Figure~\ref{fig:density-diff} shows the estimated distribution of $\Delta \alpha_{\rm ml}$ with the identification of the expected $1 \sigma$ and $2 \sigma$ quantiles. It appears that a fluctuation of $\Delta \alpha_{\rm ml} \pm 0.3$ is expected at $1 \sigma$ levels, implying that about 32\% of the systems with true common $\alpha_{\rm ml}$ values can be reconstructed with differences higher than this only owing to the observational errors.  

This results should be carefully considered because a difference of 0.3 in $\alpha_{\rm ml}$ is as high as 15\% of the solar-scaled value. For investigations that report a difference in mixing-length values of two stars in a binary system lower than this one should consider the possibility that a random error on the observables can indeed explain this discrepancy.

\subsection{Fully coupled recovery}\label{sec:fulljoint}

The last explored reconstruction also imposes the constraint of a common mixing-length value for the two stars, beside those on the common original chemical composition and age, discussed in Sect.~\ref{sec:partially-dep}. In nature, a similar assumption is theoretically justified when stars have a similar mass and are in the same evolutionary phase, but can be otherwise questioned. For our mock data this assumption is not valid for most of the systems, which are sampled with different mixing-length values for the two stars.  
The assumption of a common $\alpha_{\rm ml}$ is thus useful as a test of the robustness of the fit to wrong assumptions in the mixing-length values. 

\begin{figure*}
	\centering
	\includegraphics[height=16.5cm,angle=-90]{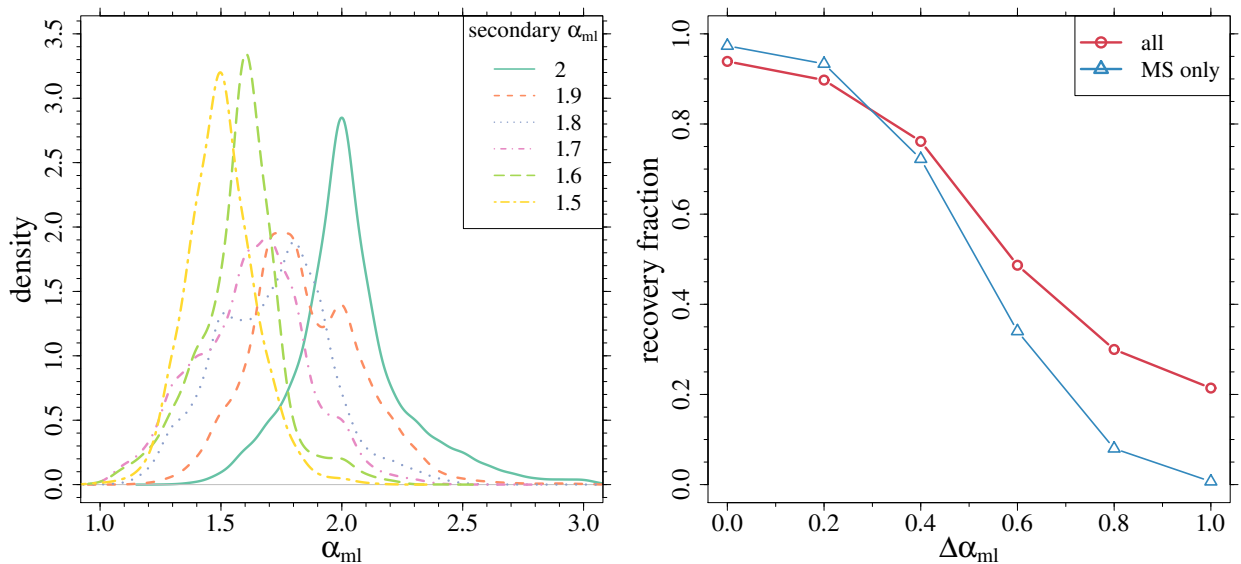}
	\caption{({\bf Left}): density of probability of the recovered mixing-length value when a common $\alpha_{\rm ml}$ is imposed in addition to the constraint of common age and original chemical composition. ({\bf Right}): fraction of systems with a valid mixing-length value fit. The 
		circles identify all the solutions, while triangles correspond only to MS cases.}
	\label{fig:alpha-uniche}
\end{figure*}

As shown in Fig.~\ref{fig:alpha-uniche} and in Table~\ref{tab:coupled} the reconstructed mixing-length values cluster around the value of the secondary star, with a low variability. This phenomenon is easily comprehensible in the light of the discussion in Sect.~\ref{sec:partially-dep}. The posterior density of the secondary is much more peaked than that of the primary, thus providing a much stronger constraint on the joint estimate. 

In the extreme scenario of $\Delta \alpha_{\rm ml} = 1.0$ the mixing-length value of the primary is underestimated from 2.5 to 1.5. This severe underestimation comes at one cost: only a marginal share of systems in this configuration can be reconstructed by the algorithm. For the vast majority the fitting pipeline is not able to provide an acceptable fit, thus suggesting the existence of some wrong assumptions in the modelling. Indeed, Fig.~\ref{fig:alpha-uniche}B shows the fraction of systems for which a fit was possible. While a common mixing-length value is found for more than three quarter of systems with $\Delta \alpha_{\rm ml} \leq 0.4$, this fraction rapidly drops to 20\% for $\Delta \alpha_{\rm ml} = 1.0$. Moreover, restricting the analysis to systems with both stars in the MS, the fraction of systems for which the fit was possible decreases even more and is nearly zero for $\Delta \alpha_{\rm ml} = 1.0$.   

Therefore it seems that large discrepancies between the $\alpha_{\rm ml}$ assumptions in the fit and in the mock data are easily detected. This is not the case for moderate differences: in this case the fitting algorithm is able to provide a common solution, which is however biased towards $\alpha_{\rm ml,2}$.

\begin{table}[ht]
 \caption{Means and standard deviations of the recovered mixing-length values in the full coupled scenario, in dependence on the $\Delta \alpha_{\rm ml}$ adopted in the generation.} 
 \label{tab:coupled}
	\centering
	\begin{tabular}{lll|cc}
		\hline\hline
		$\Delta \alpha_{\rm ml}$ & $\alpha_{\rm ml,1}$ & $\alpha_{\rm ml,2}$ &Mean & Sd \\ 
		\hline
		0.0 & 2.0 & 2.0 & 2.03 & 0.23 \\ 
		0.2 & 2.1 & 1.9 & 1.85 & 0.23 \\ 
		0.4 & 2.2 & 1.8 & 1.72 & 0.23 \\ 
		0.6 & 2.3 & 1.7 & 1.63 & 0.23 \\ 
		0.8 & 2.4 & 1.6 & 1.58 & 0.17 \\ 
		1.0 & 2.5 & 1.5 & 1.50 & 0.14 \\ 
		\hline
	\end{tabular}
\end{table}

\subsection{The impact of the observational error in the effective temperature}\label{sec:teff}

The results presented in the previous sections assume a fixed error on the observational constraints. It is interesting to explore how these assumptions influence the outcome of the fit in the three explored configurations. Due to the strong dependence of the effective temperature on the adopted mixing-length we repeated the analysis assuming an error of 50 K in $T_{\rm eff}$ that is, one half of what previously assumed. 

The results are summarised in Table~\ref{tab:teff}. A comparison with Tables~\ref{tab:independent}, \ref{tab:dep-no-alpha} and \ref{tab:coupled} shows a minor impact of this change. Overall one can observe a moderate reduction of the standard deviations, and a equally small reduction of the biases in the recovered mixing-length values. 
Overall the reduction of the standard deviation of the recovered mixing-length values in the three scenarios is about 20\%, with respect to a 50\% reduction of the $T_{\rm eff}$ uncertainty.
Therefore the results and the trends discussed so far can be considered robust against this particular source of uncertainty.

\begin{table}[ht]
	\caption{Means and standard deviations of the recovered mixing-length values adopting an observational error of 50 K in $T_{\rm eff}$. The results cover all the three considered  scenarios, in dependence on the $\alpha_{\rm ml,1}$ and $\alpha_{\rm ml,2}$ adopted in the generation.} 
	\label{tab:teff}
	\centering
	\small
	\begin{tabular}{cc|cccc|cccc|cc}
		\hline\hline
		& & \multicolumn{4}{c|}{Independent} & \multicolumn{4}{c|}{Coupled} & \multicolumn{2}{c}{Fully coupled} \\
		& & \multicolumn{2}{c}{Primary} & \multicolumn{2}{c|}{Secondary} & \multicolumn{2}{c}{Primary} & \multicolumn{2}{c|}{Secondary} & &\\
		$\alpha_{\rm ml,1}$ & $\alpha_{\rm ml,2}$ & Mean & Sd & Mean & Sd & Mean & Sd & Mean & Sd & Mean & Sd \\ 
		\hline
		2.0 & 2.0 & 2.02 & 0.34 & 2.06 & 0.34 & 2.00 & 0.36 & 2.03 & 0.22 & 2.03 & 0.22 \\ 
		2.1 & 1.9 & 2.12 & 0.35 & 1.97 & 0.33 & 2.09 & 0.38 & 1.91 & 0.19 & 1.91 & 0.19 \\ 
		2.2 & 1.8 & 2.20 & 0.34 & 1.86 & 0.32 & 2.18 & 0.38 & 1.81 & 0.17 & 1.81 & 0.17 \\ 
		2.3 & 1.7 & 2.29 & 0.34 & 1.76 & 0.30 & 2.29 & 0.37 & 1.71 & 0.15 & 1.71 & 0.15 \\ 
		2.4 & 1.6 & 2.38 & 0.33 & 1.65 & 0.28 & 2.37 & 0.36 & 1.60 & 0.13 & 1.60 & 0.13 \\ 
		2.5 & 1.5 & 2.45 & 0.32 & 1.54 & 0.26 & 2.46 & 0.35 & 1.49 & 0.11 & 1.49 & 0.11 \\ 
		\hline
	\end{tabular}
\end{table}

\section{Conclusions}\label{sec:conclusions}

We performed a theoretical investigation on the biases and random uncertainties affecting the calibration of the mixing-length value from a mock eclipsing double-lined binary system, composed by a primary artificial star of mass $M_1 = 0.95$ $M_\odot$ and a secondary of mass $M_2 = 0.85$ $M_\odot$. We used the SCEPtER-binary pipeline \citep{binary} to estimate the mixing-length of the mock stars, adopting as observational constraint the effective temperature, the metallicity [Fe/H], the radius, and the mass of the two stars. The comparison between the true and the estimated mixing-length values allows to evaluate the calibration reliability.

More in detail, several Monte Carlo simulations were conducted considering nine different scenarios, consisting on three metallicities, coupled to three different evolutionary stages of the primary (0.6, 0.9 and 1.2 of the central hydrogen exhaustion time). For each configuration, data were sampled assuming an increasing difference between the mixing-length of the two stars, from perfect agreement at $\alpha_{\rm ml,1} = \alpha_{\rm ml,2} = 2.0$ to a maximum difference of 1.0 ($\alpha_{\rm ml,1} = 2.5$, $\alpha_{\rm ml,2} = 1.5$).

The mixing length values were then estimated adopting different hypothesis in the recovery procedure. In the first case we assumed full independence between the two stars and reconstructed them without imposing any constraint in age and chemical composition between the stars. A great difficulty to estimate the mixing-length value for the two stars resulted under these hypotheses. The standard deviation of the recovered values was about 0.40, confirming the difficulties pointed out for field stars in \citet{ML-campo}. Thus, even from stars with exceptionally well constrained masses and radii, the calibration of the mixing-length parameter seems unreliable.
 
In the second case we imposed the constraint of common age and initial chemical composition for the two stars in the recovery. While the fitted $\alpha_{\rm ml,1}$ values closely match those  recovered under full independence,  the $\alpha_{\rm ml,2}$ values are much more peaked around  unbiased estimates of the values adopted in the sampling. The standard deviations of the recovered $\alpha_{\rm ml,2}$ values range from one half to one third of those of $\alpha_{\rm ml,1}$. This occurs because the primary star provides a much more tight age constraint in the joint fit than the secondary. This leads to the rejection of several extreme solutions for the secondary, while the solutions of the primary are unaffected. 
In this scenario we also explored the difference in the mixing-length values of the two stars due to random fluctuations owing the observational errors. We considered stars sampled at common $\alpha_{\rm ml} = 2.0$ and focussed the analysis on the distribution of the differences $\alpha_{\rm ml,1} - \alpha_{\rm ml,2}$. We found that the posterior distribution of these differences was peaked around zero, with a somewhat large standard deviation of 0.3 (about 15\% of the solar-scaled value). Therefore about 32\% of systems with true identical $\alpha_{\rm ml}$ are expected to show differences higher than that caused only by random errors. This results should be carefully considered when obtaining a fit from a real binary system, because a difference lower than this one has a great chance to be only a random fluctuation.

In the third case we also imposed the constraint of common mixing-length value for the two stars, beside those on chemical composition and age. Two interesting effects were detected. First, the estimated common mixing-length is mainly dictated by the value of $\alpha_{\rm ml,2}$. This happens because, as discussed above, the posterior distribution of the mixing-length of the secondary star under partial independence is much more peaked than that of the primary, thus dominating in the joint estimate process. Second, an increasing share of systems cannot be fitted by the algorithm as the differences between the true $\alpha_{\rm ml}$ increase. For $\Delta \alpha_{\rm ml} > 0.4$ less than half of the systems can be recovered; at $\Delta \alpha_{\rm ml} = 1.0$ the values decreases at 20\% . Therefore it seems that moderate differences in the mixing-length value between the two stars are difficult to detect and in these cases the solution is biased towards $\alpha_{\rm ml,2}$.

While most of the results presented in this paper can be considered general, such as the effect of the shrink of the estimated mixing-length value around the value of the secondary star,
nonetheless this work deals only with a specific binary system with fixed masses. Therefore the quantitative results cannot be expected to be valid without modifications for other binary systems with different masses and in different evolutionary phases. 
Although to check the robustness of the results presented here for different ranges of mass or by adopting a different mass ratio would be highly desirable, this possibility is actually limited by the very huge computational burden required to compute the stellar models for the recovery at the required level of accuracy. For these reasons our study should be considered as a first step in this exploration,adopting a quite common value of masses and mass ratio.
More theoretical investigations are required to fully address this open topic.

\section*{Author Contributions}

GV and MDO computed the stellar models grid and performed the recovery. GV, MDO, PGPM, and SDI wrote the manuscript.

\section*{Funding}
This work has been supported by PRA Universit\`{a} di Pisa 2018-2019 
(\emph{Le stelle come laboratori cosmici di Fisica fondamentale}, PI: S. Degl'Innocenti) and by INFN (\emph{Iniziativa specifica TAsP}).

\bibliographystyle{aa}
\bibliography{biblio}

\end{document}